# Does the Lunar Surface Still Offer Value As a Site for Astronomical Observatories?

Daniel F. Lester[1,4], Harold W. Yorke[2], John C. Mather[3]
[1]Department of Astronomy and McDonald Observatory, University of Texas, Austin TX 78712
[2]Division of Earth and Space Science, Jet Propulsion Laboratory, Pasadena CA 91109
[3]Lab for Astronomy and Solar Physics, Goddard Space Flight Center, Greenbelt MD 20771

**Abstract:** Current thinking about the Moon as a destination has revitalized interest in lunar astronomical observatories. Once seen by a large scientific community as a highly enabling site, the dramatic improvement in capabilities for free-space observatories prompts reevaluation of this interest. Whereas the lunar surface offers huge performance advantages for astronomy over terrestrial sites, free-space locales such as Earth orbit or Lagrange points offer performance that is superior to what could be achieved on the Moon. While astronomy from the Moon may be cost effective once infrastructure is there, it is in many respects no longer clearly enabling compared to free space.

**I. Introduction**

There is a rich history of thought and creativity about opportunities for astronomical observatories on the surface of the Moon. In a visionary paper written almost forty years ago, Tifft [1] considered the prospects. While the Moon is the target of few currently funded space efforts, explorations of that body, by both humans and robotic landers, represent some of the most profound achievements of our culture. In situ studies of the surface of the Moon itself are fundamentally important scientific drivers for lunar exploration, and have recently been strongly endorsed by the science community as forthcoming NASA New Horizons missions.

Many articles have since been published and talks given on specific observatory concepts, but excellent general site-specific summaries have been provided by Burns [2], Foing [3], and Hilchey and Nein [4]. Recent papers that are somewhat more focused were presented at the 2003 International Lunar Conference. Definitive compendia from earlier dates were produced by the American Institute of Physics [5, 6] and in references therein.

As a site for telescopes, we have a fairly good understanding of the conditions, opportunities, and challenges that would be involved on the Moon. Compared to terrestrial sites for astronomical observatories, the lunar environment offers enormous scientific advantages. Recognizing that the construction of lunar observatories by humans based there implies substantial incremental costs and is likely to impact mission schedules, however, the question of whether the potential benefits of a lunar telescope are justified has to be addressed. Assuming that there are other good reasons for returning to the Moon and perhaps developing a lunar base, should we add one or more observatories to the list of facilities to be built or maintained? By analogy, astronomers now enjoy

---

[4]Tel: +1-512-471-3442 ; E-mail address: dfl@astro.as.utexas.edu  (D. F. Lester)

benefits of routine science operations at the South Pole, in which other national commitments substantially offset infrastructure and transportation costs. The South Pole offers an enabling site for astronomy at a relatively modest incremental cost. Would the same hold true for a lunar observatory?

In a contemporary popular commentary for Mercury magazine, NASA geologist Paul Lowman [7] suggests that recent technological advances such as large ultra-lightweight optics, new composites for strong lightweight structures, and capable telerobotics have now made the deployment and operation of highly autonomous lunar observatories cost effective. Whether observing is done remotely by an Earth-based astronomer or more robotically by an observatory with some intrinsic intelligence, Lowman poses, as his title, the provocative question "Is the Moon the next logical location for an observatory?" He suggests that high scientific priority astronomical investments currently envisioned as free flyers may well belong (scientifically and economically) on the surface of the Moon.

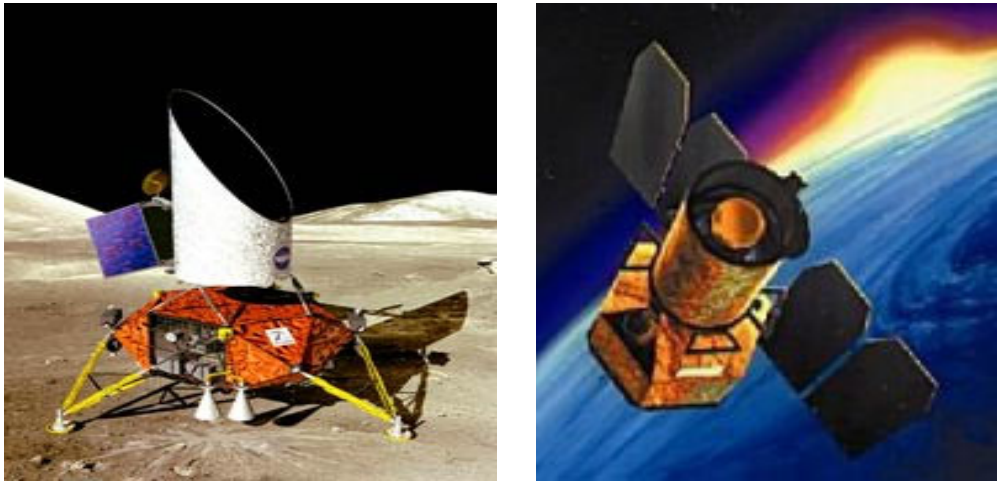

**Figure 1:** Artist's rendering of **(left)** the Lunar Ultraviolet Telescope Experiment (LUTE) concept circa-1993, and **(right)** the Galaxy Evolution Explorer (GALEX). GALEX was successfully launched in April 2003 as part of the NASA Small Explorer (SMEX) program. Both were designed to operate autonomously as ultraviolet survey telescopes, offering similar light collecting aperture and spatial resolution, roughly comparable deployed mass and power budget, with design lifetimes of two years. LUTE was conceived as a small-field transit instrument operating on the lunar surface, while GALEX is an all-sky survey mission in LEO. At the time that GALEX was competitively selected by peer review, the SMEX program had a cost cap of $120M per mission including development, launch, mission operations, and data analysis. Now successfully implemented, GALEX has been a highly economical approach to ultraviolet survey astronomy.

## II. Reexamination of Lunar-Based Astronomy: Background

In the spirit of promoting dialog on the subject, we reexamine the arguments for lunar astronomy in the light of these recent technological advances.[1] In addition to their positive impact on the feasibility of lunar observatories, these and other advances offer similar advantages to observatories in free space, whether in Earth orbit or elsewhere. We contend that while there are astronomical specialties that would enjoy undeniable advantages from the surface of the Moon, the expertise that we have gained in designing, deploying, servicing, and operating free flying space telescopes is truly remarkable, and comparatively enabling. Since this expertise now in hand was unavailable and difficult to anticipate at the time that humans were actually drilling holes, collecting rocks, and hitting golf balls on the lunar surface, early enthusiasm for lunar-based astronomy developed accordingly. For example, a Lunar Outpost program was endorsed as a long-range goal for astronomy in the Working Papers [8] for the 1990 NRC Decadal (Bahcall) Report [9]. Much of this enthusiasm was well before the Hubble Space Telescope (HST)[2], for example, and roughly contemporaneous with the Orbiting Astronomical Observatory (OAO-2), which was in many respects the first space telescope. While OAO-2 was scientifically revolutionary, its capabilities would now be considered primitive. As a result of this new expertise, we argue that while the Moon could well be a logical steppingstone for the expansion of mankind through the solar system, and the establishment of human-tended bases on the Moon may thereby enable astronomical facilities of great importance, it no longer offers clear advantages as an observatory site for what are now considered our most important astronomy goals. In many respects for observatories, the surface of the Moon is substantially inferior to free space – in low Earth orbit (LEO), Sun-Earth L2 (EL2), or elsewhere. This may be even more the case for future non-astronomical large imaging systems, a topic better discussed elsewhere.

These important astronomy goals are formally established every decade as a result of community deliberation and consensus [10]. The power of such National Academy strategic planning is paralleled by that organized triennially by the NASA Office of Space Science. Alexander [11] has underscored the political and programmatic importance of such efforts in these pages. Such goals have not intentionally sidestepped human involvement in space observatories, as exemplified by the hugely successful HST program or by even more ambitious recommendations that never came to pass, such as the Large Deployable Reflector (LDR). Such strategic planning efforts constitute the forum in which the importance of lunar-based astronomy to the astronomical community would be expressed. While consideration of missions to the Moon to make in-situ studies were specifically omitted from the charge to the Academy astronomy decadal survey committee, deployment of autonomous telescopes there to do astronomy looking outward was not, and we believe the absence of such missions from the resulting priority lists in the most recent survey is defensible [8].

We furthermore believe it is important that any reexamination of arguments for lunar-based astronomy be done with strong involvement by members of the astronomical community with an eye to scientific value, rather than by established lunar development advocates. The question is not whether it is possible to do broadly based and important astronomy from the Moon (it certainly is), or whether doing so helps build a broader space exploration initiative, but whether the scientific promise of lunar surface siting justifies the investment in both dollar cost and risk compared to other locations, and whether that scientific promise is truly enabling. Astronomy is perhaps something that we will do once we are back on the Moon, but is it really a reason to go there?

**III. Do Potential Advantages Still Apply?**

The potential advantages of lunar bases for astronomical telescopes have been discussed by many authors (see references above), and we address these advantages individually here. While the lack of atmosphere (opacity, thermal conductivity, and wind) offers stunning advantage of lunar sites over terrestrial sites, that advantage is clearly shared with free space. Uninterrupted studies of the Earth itself have been cited as one reason for lunar observatories. While such work is only peripherally relevant to our discussion of astronomy here, we believe that such a site has few performance advantages for Earth science, if any, over low or geosynchronous Earth orbit or at Sun-Earth L1 (EL1), a semi-stable location where the Earth is seen continuously illuminated. This location is now used for solar surveillance and is a scientifically endorsed future site for the Triana Earth-observing mission.

**Lack of magnetic field:** Unlike the Earth, the Moon has little or no magnetic field, and as a result has no radiation belts to trap and concentrate plasma, largely from the solar wind. The charged particle flux there, and in cis-lunar space in general, is thus substantially lower than in high Earth orbits that intersect these radiation belts. Below the belts, such as in LEO, the Earth's field offers effective shielding for spacecraft. Such particles produce effects in sensor arrays, and the longest exposures even from LEO always show hot pixels, and slow degradation of the detectors as a result of particle impacts and implantations. The lunar surface can, in principle, offer complete shielding toward half the sky from solar wind particles. While noticeable, annoying effects in free space are routinely mitigated by on-board shielding, which is effective against low energy solar wind plasma. Proper operations management (e.g. Chandra in high orbit, HST in LEO) is of value too, in scheduling observations optimally.

**Lack of residual atmosphere:** Lunar sites are not affected by the high altitude residual atmosphere that spacecraft encounter in LEO. These molecules can have long-term deleterious effects on optical coatings that are exposed to them. Also, residual $O_2$ impacting structures and outgassing products at orbital velocity (~8 km/s) excites faint emission that can contaminate the most sensitive astronomical observations. These effects are rendered largely harmless, however, by pointing orbiting telescopes "downwind", and by including minimal shielding (e.g. HST).

**Lack of orbital debris:** LEO is getting to be an increasingly dirty place, with debris on all size scales from generations of use and misuse. The Moon has no such debris of human origin. Whereas there is little hope for recovering an observatory that is hit at high velocity by even a marble-sized chunk of this debris, the threat level in LEO is not high. Damage to HST, for example, has been minimal, with the equivalent of about a half-dozen sand-grain sized hits per year that each leave a tiny dent in the housing. An assessment of orbital debris for the National Academies [12] derived an estimated impact risk of 0.1% per year for a >1cm piece of debris on a 10 $m^2$ satellite cross section in LEO, with the risk biased strongly to orbits significantly higher than the International Space Station (ISS) or HST. Real risk to astronomical observatories in Earth orbit has, thus far, not been demonstrably significant.

**Stable thermal environment:** As a result of the slow motion of the Sun across the sky, telescopes on the Moon have more time to thermally equilibrate than do telescopes in LEO at the low orbital inclination that allows maximally efficient deployment. The thermally induced expansion and contraction of spacecraft in LEO produce image motion which, if not compensated for, results in target tracking errors. The effect is small for well designed and properly shielded optical systems, however, and the timescale is well within the bandwidth of star trackers. For infrared telescopes, thermal equilibration is important for optimal sensitivity, though proper shielding and attention to potential scattering paths has made this effect unimportant in LEO cryogenic telescopes, e.g. the Cosmic Background Explorer (COBE), the Infrared Astronomical Satellite (IRAS), and the future Wide Field Infrared Survey Explorer (WISE), all of which use polar sun-synchronous orbits over the terminator.

The four lunar properties listed above, while somewhat advantageous in comparison to observatory sites in Earth orbit, are unexceptional when compared to more remote sites such as the Earth's Lagrange points (e.g. the Wilkinson Microwave Anisotropy Probe – WMAP at EL2, the Solar and Heliospheric Observatory – SOHO and Genesis at EL1) and drift-away heliocentric orbits (e.g. the Spitzer Space Telescope – SST, formerly known as SIRTF).

**Solid surface:** Lowman states in his Mercury article that the briefest answer to the question about scientific advantages of the lunar surface is the word "surface." The lunar surface offers a stable foundation, a reaction load and inertial sink to push against for movement, and a fixed reference frame. This was of principal value for early concepts of space telescopes that were by necessity human-attended. Such telescopes (e.g. Tifft [1]), in which astronomical operations involved changing of photographic plates between exposures that were guided by a human eye, saw induced vibration from human beings in the living and working quarters as being a major technical impediment. The earliest wisdom about telescopes was thus that one needed a good solid pier (ideally with concrete anchored in bedrock!) to mount a telescope properly, and the lunar surface offers this opportunity. While this rule-of-thumb wisdom has guided our efforts for generations of terrestrial telescopes, it is now understood to be unimportant (e.g. HST and many military surveillance systems) for free space.

Free space is actually a very stable place to put a telescope. If you don't push on it, it won't turn. In fact, the forces that exert torques on large structures in the vicinity of the Earth are weak, and we have achieved real mastery of precision pointing through a combination of gyro control and star tracking at levels of precision that greatly exceed that of terrestrial telescopes. Such systems are no longer considered particularly challenging, and attitude control system failure has been the cause of death of just a few orbiting astronomy instruments: The Advanced Satellite for Cosmology and Astrophysics (ASCA) was well into it's extended mission in a reduced orbit when the control system was unable to compensate for heating of the Earth's atmosphere by a solar storm. The Tomographic Experiment using Radiative Recombinative Ionospheric EUV and Radio Sources (TERRIERS) was rendered unusable because of a software problem, and the Solar Maximum Mission (SMM) failed three years before it was repaired by astronauts. Star trackers and drive mechanisms on a lunar surface telescope can fail similarly. (Fixed attitude telescopes without tracking mechanisms have been proposed for implementation on the Moon, and these will be addressed below.) For a telescope in free space, pointing is achieved by using magnetic fields, momentum wheels, and jets to reorient under servo control using star trackers and gyros for sensing. This is done without large drive mechanisms or loaded bearings that could be problematic for a telescope that needs to push smoothly and reliably against a reaction load (the Moon).

The solid lunar surface does offer a kind of huge vacuum optical bench for large astronomical interferometers involving arrays of telescopes. Once the interferometer elements are optically coupled their baseline is relatively secure as the Moon, without tectonic action, provides an acoustically and vibrationally quiet platform. Autonomous deployment of such an array on the lunar surface, perhaps involving several discrete precision landings, presents a huge challenge however. For space interferometers with modest baselines (e.g. the Space Interferometry Mission – SIM), an integral truss allows for changes in the baseline scale size and, even more simply, changes in position angle that offer complete UV plane coverage. Such options are vastly more difficult with an array of linked telescopes resting on the lunar surface, requiring mobile structures and agile phase delay lines to synthesize that coverage.

For large baseline free-space interferometers, in which an integral truss is impractical, individual free-flying telescopes will have to do precision formation flying and accurate fringe tracking. While such formation flying has not yet been achieved yet in space, technology demonstrators (e.g. the Starlight effort, formerly ST-3 in the New Millennium Program) have given us confidence in our ability to do it reliably. As a result of this confidence, such precision formation flying is already specified for the Laser Interferometer Space Antenna (LISA) gravitational wave observatory for which development has now begun, with a planned launch in 2011. For this near-term mission, relative positioning knowledge at a 10 nm level over the projected $5x10^6$ km baseline are required, with actual spacecraft positioning being maintained at the centimeter level. The optical interferometer envisioned for the flagship mission Terrestrial Planet Finder (TPF) would rely on spacecraft positioning precision a factor of ten better, in order to allow accurate fringe-following. In-space demonstration missions under the New Millennium program are being considered for this technology. Such formation flying also allows for reorientation of the entire array to point anywhere on the celestial sphere with the same UV-plane coverage, which a fixed lunar array would not be able to do. Formation flying does present added risks, however, such as the potential for collisions. In addition, gravity gradients will tend to slowly pull the formation apart. But these effects should be manageable by careful design and operations planning.

In summary, the solid surface of the Moon has been advanced as a quality that is technologically enabling for astronomy in the absence of expertise in pointing, tracking, acquisition and alignment of free-space observatories. We believe that this expertise is now largely in hand, and the reasoning that favors solid surface operations is in most cases no longer compelling.

**Gravity:** Although there are many reasons to believe that zero-g conditions actually make it easier to deploy and operate a large telescope, arguments have been presented that small amounts of gravity (like the 1/6 g found on Moon) can be helpful. For human-aided efforts this is understandable, because things you let go of end up at your feet, and specialized tools and restraints may not be needed. In addition, dust kicked up by surface operations ends up back on the ground. However, gravity does present loading problems with concomitant structural deformations that are direction-dependent. Where you look in the sky determines how your telescope is aligned. The telescope must be stiffer and heavier in order to work correctly. For autonomous telescopes that require packaging into finite sized shrouds and deployment out of them, extra stiffness comes at a high price, including poorer thermal isolation. Finally, it must be stated that gravity presents, to telescopes headed to the Moon from the Earth, the single largest risk factor. Managing it means fighting it, at least on the way down to the lunar surface. It is worth underscoring here that one of the major issues for the planning of extraterrestrial telescopes is whether gravity is an advantage or a disadvantage.

In this context, a new concept for a lunar observatory [13] has been advanced by Roger Angel of the University of Arizona that actually depends on gravity, and is very much otherwise enabled by an extraterrestrial site. While it is not intended to be autonomously deployed, its innovative nature deserves mention here. Angel proposes a large parabolic primary mirror generated by rotating a thin circular tray of fluid. This kind of telescope has been proven terrestrially with liquid mercury [14], which also offers a specular surface. Angel proposes a mirror material that is liquid at 90 K, with a flashed-on reflective coating. While such a telescope is intended for use only in zenith surveys, as it only points along the gravity vector, it allows a huge collecting area to be created with what is in principle relatively simple equipment. The Moon offers such a telescope not only the gravity that is needed to generate the figure, but an environment entirely free of the wind that, by introducing ripples in the surface, limits its terrestrial usefulness. Such a telescope could not be constructed in free space and in principle offers an aperture size that could only otherwise be created by large arrays of individual telescopes. Serious questions remain about this concept, however, which requires detailed design work and technical review. Questions include the suitability of materials, including the liquid mirror, power systems, construction and deployment, contamination, and the details of the local lunar environment.

**Slow sidereal rate:** At any given site on the Moon, the celestial sphere appears to rotate over the gravity vector about thirty times more slowly than it does on the Earth. Lunar astronomy advocates have proposed telescopes that use this slow motion advantageously. Transit telescopes that survey a strip of sky by watching it move slowly overhead allow in principle for the simplest possible telescope – in which no guiding or tracking is required, and the telescope need not even be pointable. Modest integration times to see faint astronomical sources can be achieved by clocking the CCD imager readout at a rate that matches the slow sidereal rate. Such strip surveys have value in population censuses of the universe, and such tracking strategy has been achieved (though with much shorter exposure times) with some success in terrestrial testbeds, such as the Charge Transfer Instrument (CTI) [15]. Autonomously deployed lunar reference designs, e.g. the Lunar Ultraviolet Telescope Experiment (LUTE – see figure 1), and the Lunar Transit Telescope (LTT) have been examined in this context, and 1 m-class aperture zenith survey telescopes there have been found to be feasible [16]. For more capable lunar telescopes that can track and guide across the sky the site offers, in principle, uninterrupted integration times of order a week.

In practice, such lunar opportunities have never been shown to offer scientific benefits unobtainable by other means. Terrestrial survey telescopes (e.g. the Sloan Digital Sky Survey - SDSS) now achieve optical detection performance similar to that specified for the lunar precursor telescopes. As noted above, tracking and guiding from free space is no longer an extremely difficult task. The Galaxy Evolution Explorer (GALEX) mission, now in LEO (see Figure 1), is in the process of mapping the ultraviolet sky to a depth similar to what LUTE would have done, and is a highly economical approach to that science. The advantage of long <u>uninterrupted</u> integration times is, with the advent of ultra-low readout noise sensor arrays, and in the presence of cosmic rays, no longer of great importance. To the extent that the science goals justify ultra-long exposures, adding multiple shorter exposures has been shown to be a useful strategy, and the penalty for reacquisition, should that be necessary, is not severe. For broadband survey imaging, such long exposures with diffraction-limited pixels would be overwhelmingly zodiacal background limited anyway. We note that the Supernova Acceleration Probe (SNAP) concept for a 2 m aperture optical and near infrared deep survey telescope in LEO is in many respects the scientific descendent of lunar survey telescopes that have been previously proposed. SNAP combines fairly mature telescope technology and detector concepts and, aside from the huge focal plane arrays that are envisioned for it, is a relatively economical observatory.

**Radio quiet:** Some of the earliest concepts of lunar astronomy have focused on the radio isolation of the lunar far side. Radio telescopes there would be effectively shielded from terrestrial interference both manmade and natural, the latter noise from magnetically accelerated charged particles at the poles of the Earth. The shielded zone of the moon (SZM) has been formally identified in the International Telecommunication Union (ITU) Radio Regulations as being worthy of protection. The science value of these centimeter-meter scale wavelengths has been referenced in community strategic planning for wavelengths accessible from the ground, e.g. the Square Kilometer Array (SKA) and the Low Frequency Array (LOFAR), though without assigned priority. Low cost (MIDEX-scale) free-space arrays to do decametric radio interferometry have already been proposed (e.g. the Astronomical Low Frequency Array – ALFA, and the Solar Imaging Radio Array – SIRA) as these wavelengths are blocked entirely by the ionosphere. Such arrays would be on station at distances on a scale of that of the Moon where terrestrial radio interference is quite low, and where performance would be limited by cosmic backgrounds. It should be noted that modern technology allows powerful approaches to radio interference mitigation in any case, and also that no near-Earth site can be expected to remain completely radio quiet as the solar system is explored, for example with an aggressive development of Mars [17].

**Cold:** Both for functionality and low noise operation of astronomical sensors, and for reduction in thermal background radiation for infrared sensors in particular, lowering the temperature of astronomical observatories to near absolute zero has been a holy grail, and as a result of atmospheric condensation, is impossible to do with terrestrial telescopes. While reflective shielding allows space observatories to be passively cooled, allowing on-board cryogens (if needed) to be used to more efficiently chill the systems, residual heat loading from the Sun and the Earth itself make expendable cryogen lifetime a serious mission design issue. Space-qualified mechanical cryocoolers with sufficient cooling capacity and adequate reliability are just now being developed, and are not particularly efficient in their use of electrical power. With either expendable cryogens or cryocoolers, observatory thermal management is a costly design issue. In this respect, the Moon offers perhaps its most important potential astronomical advantage over low Earth orbit, at least in the long term.

The orbit of the Moon is not far out of the ecliptic, and its rotational axis is similarly aligned. As a result, the poles of the Moon see the Sun and Earth only very close to the horizon. At the bottom of craters near the lunar poles, the Sun and Earth never rise. The heat load there is dominated by scattered and diffracted light from surrounding mountain peaks, and small amounts of interior heat conducted to the surface. Although temperatures have never been measured there, it has been estimated that the surface could naturally and perennially be as cold as 30 K. In practice, a crater floor at the pole of the Moon may be the coldest place in the entire inner solar system. With good shielding, telescopes there could passively reach temperatures around 7 K, heated by the light from stars and interplanetary dust. At such low temperatures, even mirrors contaminated with small amounts of lunar dust would perform adequately for thermal infrared measurements. Superconducting bearings for support are conveniently managed. We note that while non-polar lunar sites do get cool when the Sun is down, thermal inertia of the surface leads to nighttime temperatures that are not difficult to achieve passively in free space, even in LEO.

The use of the lunar poles for infrared telescopes was broadly sketched out by Lester [18, 19] after discussions with the insightful and creative Harlan Smith, whose leadership in this subject has been impressive (see Smith [20]). As a site for infrared telescopes, the Moon offers some remarkable properties. Concerted design efforts along these lines have been made by van Susante, Duke, and their collaborators under NASA's Revolutionary Aerospace Concepts program [21], in a recent effort partially supported by NASA's Office of Space Science. In addition, Roger Angel's liquid mirror telescope described above specifically targets the lunar pole as a possible site.

But observatory operations at a lunar pole would not be easy. While the topography has been mapped at low resolution by radar studies from Earth, the coldest polar regions have not been surveyed in detail optically (either from the Earth or from lunar orbit) because they are not well illuminated. They are so cold, in fact, that they are not even surveyable by their own thermal emission. Consequently, plans for such lunar polar telescopes require substantial survey work, perhaps from fly-over radar studies, low-light imaging, or actual visits with illumination in tow. In addition, deep studies of the thermal far-infrared and submillimeter sky with cold telescopes are expected to be, as a result of diffraction limited beam sizes, strongly confusion limited. As a result, small telescopes (of 1-2 m sizes comparable to that of lunar concepts LTT and LUTE) are of limited value for far-infrared work, and large cold telescopes on the Moon are not at all conducive to autonomous deployment. The lack of direct sunlight means that the cold observatory will need to be powered by a remote solar collector or thermal generator, perhaps on the crater rim, whether by cable or by beamed microwaves. A similar arrangement may be needed for communication with the Earth. These design elements add increased complexity and risk for autonomous deployment and operations. Finally, sky coverage in a deep lunar polar crater will be quite limited. No such handicaps affect EL2, for example, which is the target site for the James Webb Space Telescope (JWST), Herschel, Planck, and the Single Aperture Far Infrared telescope (SAFIR), among many other missions.

Even with astronauts available to assemble a large telescope and tie it to a power source and communications link, a polar site is hardly convenient for humans. It is almost completely dark, and very cold, comparable to Pluto. For purposes of repair and reinstrumentation by astronauts it can thus be considered fairly inaccessible, and contamination of the telescope by warm gas expelled from space suits, outgassing of equipment, as well as gases that are released from surface cold traps by on-site activities, can be considered a major performance risk. Reservoirs of ice and hydrocarbons in these craters may be helpful as resources for lunar colonization, but they pose trouble for cryogenic observatories. Concepts for a mobile polar telescope have been advanced – one that it could be pulled or robotically driven in and out of the permanently shadowed crater to allow it to be serviced – but this entails much greater complexity and cost.

The infrared astronomical community looks forward to a time when lunar polar operations may be feasible. The thermal properties of the site show, in principle, obvious potential. But we are in no position at the present time to take advantage of this remarkable natural resource. Large deployable cold telescopes are now being actively planned for EL2, and the additional complexity in shielding, active cooling and thermal management would, compared with lunar polar basing, seem to offset many of the challenges that will have to be met. Dramatic improvements in development of highly efficient deployable sunshields render infrared telescopes highly enabled at EL2. The thermal shield for the James Webb Space Telescope (JWST), for example, cuts heating from the Sun and Earth by a factor of about $10^7$, such that baseline instrument temperatures of 35 K are dominated by their own power dissipation. The same would be true in a lunar polar crater. It is noteworthy that lunar gravity, while small, would require support structure connecting a lunar telescope to the surface that is substantially stronger than that needed to connect a free-flying telescope at EL2 to its shield. Such extra strength compromises the thermal insulation between the lunar surface and a telescope mounted to it.

**Lifetime:** A recurring theme in lunar-based astronomy advocacy is observatory lifetime. With cold lunar regions that may obviate the need for expendable cryogens or active cooling, and a solid surface that obviates the need for expendable station-keeping and gyro-offloading propellants, a telescope built on the Moon might be expected to last a long time. But is this good? Does a long lifetime for a state-of-art instrument have real value? This is a fundamental question of space science policy, and simple answers must be carefully qualified.

A basic precept for NASA space science strategic planning is that missions are designed to answer questions, and they are designed to do so on a scientifically valuable timescale. Telescope technology, and focal plane sensors in particular, are advancing rapidly – for the same thin-film technology reasons that result in Moore's Law. Over the last thirty years the state-of-art format size of array sensors has been doubling approximately every eighteen months. In the infrared, both format size and sensitivity have been dramatically improving, leading to an even steeper performance improvement curve. In simple terms, every eighteen months a telescope with a fixed focal plane instrument becomes half as capable as a new replacement would be. While the impact on capability is necessarily dependent on the science goal, this progress gives little incentive to keeping old instruments running for a long time. For the Hubble telescope, which has been fully functional for more than a decade, regular servicing missions involving installation of new equipment have dramatically extended its scientifically useful life. It has become a new instrument with some regularity.

Annual operations costs, which include flight operations, navigation and tracking operations, and data analysis, are substantial, and can be roughly assessed as ten percent of the capital cost of the mission (a rule of thumb based on recent Explorer missions and generic life-cycle cost models that is remarkably consistent with that for terrestrial observatories). The Office of Space Science routinely reviews extended mission plans for spacecraft, and where prudent fiscally and scientifically, will turn working missions off (e.g. the Extreme Ultraviolet Explorer – EUVE and the Eighth Interplanetary Monitoring Platform – IMP-8) in order to start new ones. In summary, the value of ultra-long lived autonomous observatories on the Moon (or elsewhere in space) is not self-evident, and will require strategic scientific justification not yet advanced.

In this paper, we have thus far approached lunar-based astronomy by reexamining advantages advanced by proponents. These advantages are noteworthy, but not compelling or unique. These same proponents have also identified potential disadvantages that are inherent to such lunar basing. Such disadvantages are commonly recognized, and need only brief mention here. Gravity, which introduces clear risks and concomitant operations costs, has been discussed above. Surface irregularities on the Moon make site operations, especially autonomous ones, challenging. Autonomous landers must be able to function in the presence of large rocks, ridges, and cracks. They must either do so with structural immunity to them that is built in, or topographical knowledge of the target site on scales smaller than the observatory, perhaps with intelligent systems to avoid obstacles. Contamination of both optics and mechanical systems by lunar dust is a matter of considerable concern. Such flying dust, which may be charged by the solar wind and securely attached to components electrostatically, arises not only as a result of landing but also from meteoroidal impacts, and from local traffic encouraged by observatories that require frequent service or upgrades. Such dust will compromise the sensitivity of telescopes that are sensitive to scattered light, such as coronagraphs, and of thermal infrared telescopes in general. It will increase wear and reduce performance on any loaded bearings. It can be assumed that careful surface operations can reduce this risk as can creative design (e.g. non-contact bearings), and thermal shielding will offer some protection against grains on ballistic trajectories. But in the context of astronaut-aided operations, it is sobering to remember that such operations on the lunar surface become severely restricted due to dust accumulation in spacesuit joints. Following the logic above, the briefest answer to the question about the scientific <u>disadvantages</u> of the lunar surface is also the word "surface"!

**Accessibility:** In view of the discussion above about lifetime, and the caveats in the discussion about lunar polar opportunities, the prospect for accessibility of a lunar observatory must be addressed. Such potential accessibility is seen as a profound advantage by lunar astronomy advocates. While this transcends the issue of autonomous deployment and operation, this long view should be considered. Whether on the Moon or in free space, it is likely that the largest and most ambitious observatory facilities will require hands-on attention from astronauts. On the basis of HST in one case, in which astronaut servicing was planned (and SMM, where such servicing was not), or the Apollo Far-Ultraviolet Camera/Spectroscope in another, astronauts have maintained astronomical facilities in both places. While such involvement of humans admits additional risks that are politically and programmatically challenging, it also reduces other risks, in principle, by making things fixable – whether by a tweak, a wholesale subsystem transplant, or a well-aimed kick (cf. unsticking the antenna on the Compton Gamma Ray Observatory – CGRO).

Given this prospect, does the Moon offer any advantages over free space for human-aided telescope construction and operations over, say, a Lagrange point such as Earth-Moon L1 (LL1), or EL2? It can be argued that while humans have never been to those latter places, they have spent a lot of time on the Moon. Recent studies by the NASA Exploration Team - NExT (H. Thronson, private communication) have, however, investigated the possibility of human tended stations at such scientifically useful Lagrange points, opening observatories based there to human intervention for purposes of deployment, repair and servicing. These Lagrange points are energetically less costly to supply and visit than bases on the lunar surface, and are recognized as potential gateways to the solar system by virtue of the low delta-V required to travel from them to many key locations. Contemporary traffic models for ES1 and ES2 show about a dozen NASA space science missions slated for these sites on a time scale of two decades. The wealth of zero-g experience that we have now developed points us more clearly toward free space than to lunar surface sites for the largest space telescopes, and such Lagrange point outposts were found by the NExT team to be entirely feasible. The programmatic viability of such outposts has been articulated cogently and most recently by lunar astronaut Edwin (Buzz) Aldrin, in a recent OpEd piece for the New York Times [22]. It is important to note, as we did above, that while human-tended astronomy from the Moon will most likely depend on other reasons for being there, the same may be said of free space, for example the Lagrange points.

**IV. Summary**

Three decades ago the United States abandoned the Moon, a decision that, from a national space policy standpoint remains highly controversial. The technological capabilities that could have encouraged subsequent astronomical operations on the Moon were never developed. Since then, several countries have methodically, strategically, and aggressively developed capabilities for deploying and operating telescopes in free space, and made huge strides in zero-g human operations. In the same way that lunar-based astronomy appeared decades ago, even in hindsight, to have the programmatic advantage, operations in free space do so now. We can now point to ambitious tasks that we have accomplished in free space, expertise that we now control, and extrapolate to the future there with some confidence. Had the decision about our continued presence on the Moon been different, much of our current free-space expertise might never have developed to its present level of sophistication and lunar-based astronomy might now be more programmatically attractive. When we actually return to the surface of the Moon, constructing bases that are continually occupied, offering infrastructure, transportation, ready service for investments there, and possibly material resources, this conclusion may be revised. In this respect, consideration needs to be given as to what critical mass of lunar habitants is necessary to even suggest construction, maintenance, and operation of a lunar telescope. But we contend that from the point of view of observatory science priorities, free-space offers important things that the lunar surface does not.

Lowman [7] concludes that "the greatest obstacle to Moon-based astronomy as a contender for available funding probably lies in its position between two now well established fields: ground-based and [free] space-based astronomy". We concur, but would add that in order to succeed, lunar astronomy advocates should present a compelling argument in which science gain, risk avoidance (both science and personnel), and overall cost (which may well include programmatic value that is long-range, as well as mission-specific) have clear advantages over observatories in free space. It is by these measures that the value of lunar based astronomy should be assessed.

**Acknowledgements**

[1] In this effort, we restrict ourselves to largely to wavelengths of light that lend themselves to image formation by focusing, longward of the hard UV. Opportunities for particle and gamma ray astrophysics will be addressed elsewhere.

[2] Here and elsewhere, descriptions of most NASA space science missions – recently planned, operating, and retired – are available at http://spacescience.nasa.gov/missions/.